\documentclass[letterpaper,journal,onecolumn]{IEEEtran}
\pdfoutput=1

\usepackage{amsmath,amsfonts}
\usepackage{algorithmic}
\usepackage{algorithm}
\usepackage{array}
\usepackage[caption=false,font=normalsize,labelfont=sf,textfont=sf]{subfig}
\usepackage{textcomp}
\usepackage{stfloats}
\usepackage{url}
\usepackage{verbatim}
\usepackage{graphicx}
\usepackage{epstopdf}
\usepackage{cite}
\usepackage{soul, color, xcolor}
\usepackage[switch]{lineno}
\usepackage{setspace}

\usepackage{balance}

\begin{document}
%\linenumbers
\title{{Blockchain-Enabled} Variational Information Bottleneck for IoT Networks}

\author{Qiong Wu,~\IEEEmembership{Senior Member,~IEEE}, Le Kuai, Pingyi Fan,~\IEEEmembership{Senior Member,~IEEE}, \\ Qiang Fan, Junhui Zhao,~\IEEEmembership{Senior Member,~IEEE}, Jiangzhou Wang,~\IEEEmembership{Fellow,~IEEE}
        % <-this % stops a space
\thanks{
This work was supported in part by the National Natural Science Foundation of China under Grant No. 61701197, in part by the open research fund of State Key Laboratory of Integrated Services Networks under Grant No. ISN23-11, in part by the National Key Research and Development Program of China under Grant No. 2021YFA1000500(4), in part by the 111 Project under Grant No. B23008. (Corresponding Author: Pingyi Fan)

Qiong Wu and Le Kuai are with the School of Internet of Things Engineering, Jiangnan University, Wuxi 214122, China, and also with the State Key Laboratory of Integrated Services Networks (Xidian University),  Xi'an 710071, China (e-mail: qiongwu@jiangnan.edu.cn, lekuai@stu.jiangnan.edu.cn)

Pingyi Fan is with the Department of Electronic Engineering, Beijing National Research Center for Information Science and Technology, Tsinghua University, Beijing 100084, China (e-mail: fpy@tsinghua.edu.cn)

Qiang Fan is with Qualcomm, San Jose, CA 95110, USA (e-mail:qf9898@gmail.com)

Junhui Zhao is with the School of Electronic and Information Engineering, Beijing Jiaotong University, Beijing 100044, China (e-mail: junhuizhao@hotmail.com).

Jiangzhou Wang is with the School of Engineering, University of Kent, CT2 7NT Canterbury, U.K. (e-mail: j.z.wang@kent.ac.uk)

}% <-this % stops a space
%\thanks{Manuscript received xx xx, 202x; revised xx xx, 202x.}}
%\thanks{Manuscript received xx xx, 202x; revised xx xx, 202x.}
}

%\markboth{Journal of \LaTeX\ Class Files,~Vol.~xx, No.~x, xx~202x}%


\maketitle

\begin{abstract}
%In Internet of Things (IoT) networks, the amount of sensed data by user devices may be huge, resulting in the serious network congestion. To solve this problem, intelligent data compression is critical. The variational information bottleneck (VIB) approach can be employed to train the encoder and decoder, so that the required transmission data size can be reduced significantly. However, VIB suffers from the computing burden and network insecurity. In this paper, we propose a blockchain-enabled VIB (BVIB) approach to relieve the computing burden while guaranteeing network security. We first construct a new network framework by splitting encoding and decoding networks to solve the computation burden and then propose a new algorithm for the framework to improve the IoT network security.
%We construct an experimental platform by integrating Python and C++ to verify the efficiency of our proposed BVIB approach. Extensive simulations demonstrate that BVIB outperforms other baseline approaches.

In Internet of Things (IoT) networks, the amount of data sensed by user devices may be huge, resulting in the serious network congestion. To solve this problem, intelligent data compression is critical. {The variational information bottleneck (VIB) approach, combined with machine learning, can be employed to train the encoder and decoder, so that the required transmission data size can be reduced significantly.} However, VIB suffers from the computing burden and network insecurity. In this paper, we propose a blockchain-enabled VIB (BVIB) approach to relieve the computing burden while guaranteeing network security. {Extensive simulations conducted by Python and C++ demonstrate that BVIB outperforms VIB by 36\%, 22\% and 57\% in terms of time and CPU cycles cost, mutual information, and accuracy under attack, respectively.}

\end{abstract}


\section{Introduction}
\subsection{Background}
With the development of Internet of Things (IoT) networks, user devices can sense various data and connect the data to networks, thus achieving more efficient information. {However, the amount of sensed data may be huge and can indeed lead to network congestion due to the limited bandwidth and processing power of the devices. In IoT networks, network congestion can lead to increased latency and even packet loss, significantly impacting the performance and reliability of the network, leading to worse performances of IoT services\cite{9064704,9785856}. To alleviate it, intelligent data compression is critical due to its ability to reduce the data size to relieve network congestion. Existing approaches to intelligent data compression in IoT networks can be divided into two categories: lossless compression, which maintains data integrity, and lossy compression, which sacrifices some data accuracy for higher compression ratios. 
The information bottleneck (IB) is a type of lossy compression that utilizes mutual information to achieve higher compression ratios\cite{tishby2000information}. }Specifically, each user device first estimates the prior probability of data, then calculates the mutual information of the data to measure the correlation among data, based on the estimated prior probability. Afterwards, it designs the codebook by maximizing the mutual information of data to generate an encoder and decoder, then adopts the encoder to encode data and restores the most useful portion of the data through decoder, which results in significantly smaller data size and thus helps alleviate network congestion. However, maximizing mutual information incurs a high mathematical complexity, which poses challenges to codebook design\cite{10313285}. Moreover, the estimation of prior probability relies on unobserved prior knowledge, which poses a challenge for the calculation of mutual information.

Deep neural networks can train encoder and decoder models instead of relying on manual codebook design, which leads to the deep information bottleneck (DIB) approach \cite{tishby2015deep}. However, the challenge regarding mutual information calculation is still unsolved. Alternatively, the variational information bottleneck (VIB) approach is further proposed based on DIB to approximate mutual information through calculating upper and lower bounds of the mutual information, which makes the challenge of DIB more tractable \cite{alemi2016deep}. However, VIB also has the following shortcomings.

\begin{itemize}
	\item Computing burden: VIB implements the encoder and decoder on the same device, but the computing resources of user devices are insufficient to support the training of encoder and decoder, which causes the computing burden for user devices. %Thus, the VIB algorithm has constraints for computation intensive applications.
%	\item Network security: The rapid development of IoT has attracted adversaries to obtain user privacy by attacking user devices, which would affect the training process and deteriorate the training accuracy\cite{wu2019trajectory}.
	\item Network security: The rapid development of IoT has attracted adversaries who launch attacks by flooding devices with a high volume of useless requests, resulting in adversely impacting device computing resources, thereby compromising training accuracy \cite{wu2019trajectory}.
\end{itemize}

%Blockchain technology deploys \hl{nodes with sufficient computing resources (commonly known as {server}s)} in IoT networks to collect data from user devices and generate the blocks including data, where each block is linked to another one to create an unbreakable chain. 
%Hence it is difficult to change the data in any block and thus the blockchain technology can guarantee the security of IoT networks. 
Blockchain technology deploys servers with sufficient computing resources in IoT networks to collect data from user devices and generate the blocks (including data, timestamp, and so on), where each block is linked to another one to create an unbreakable chain. {The consensus mechanism of blockchain technology can enhance the robustness against attack by voting or other tricks. By leveraging the consensus mechanism to exclude adversaries, blockchain technology can ensure the malicious actions do not compromise the training accuracy and guarantee the network security \cite{9868080, 10024370}. Moreover, blockchain introduces servers with sufficient computing resources which provide the feasibility to complement the insufficient computing resources of user devices by placing encoder and decoder on user devices and servers, respectively. }
Therefore, it is necessary to propose a blockchain-enabled VIB algorithm to relieve the computing burden while guaranteeing the network security for IoT networks.

\subsection{Related work}
Some VIB-based approaches have been proposed for various applications. 
%In \cite{9576573}, Xie \emph{et al.} proposed a new approach based on the multimodal extension of VIB to predict the popularity of video information. 
%In \cite{9606667}, Shao \emph{et al.} proposed a new communication framework based on VIB to reduce transmission latency. 
In \cite{10158495}, Wang \emph{et al.} combined the  self-supervised learning adopting the pretext task which is constructed by unlabeled data and VIB to alleviate the sample difficulties for the deep learning. In \cite{9813696}, Uddin \emph{et al.} proposed a loss function based on disentangled VIB approach and federated learning (FL) to reduce communication overhead. In \cite{li2023task}, Li \emph{et al.} proposed a task-oriented communication framework by using digital modulation based on VIB to achieve higher accuracy and generalization. However, all the above works have not solved the shortcomings of computing burden and network security for VIB.

Some works have considered blockchain technology to alleviate the computing burden in IoT networks. In \cite{wei2021secure}, Wei \emph{et al.} designed an obfuscating policy to shift encryption computations to reduce the computing burden in IoT networks. 
In \cite{nguyen2021secure}, Nguyen \emph{et al.} combined a smart {contract} and a double-dueling Q-network to offload the computing burden of mobile devices in IoT network with blockchain. 
In \cite{10131971}, Lan \emph{et al.} deployed unmanned aerial vehicles (UAV) in IoT network and adopted blockchain technology to alleviate the computing burden. However, these works have not considered the VIB approach to compress information for IoT networks.

To the best of our knowledge, no works have considered how to combine blockchain technology and VIB approach to relieve the computing burden while guaranteeing the network security for IoT networks, which motivates us to conduct this work.

\subsection{Contributions}
In this paper, we propose a blockchain-enabled variational information bottleneck (BVIB) approach to relieve the computing burden while guaranteeing the network security for IoT networks\footnote{The source code is available at https://github.com/qiongwu86/Blockchain-enabled-Variational-Information-Bottleneck-for-IoT-Networks}. {BVIB proposes a new framework that includes devices and servers.  The devices are deployed with encoders for extracting data. The servers are deployed with decoders while maintaining the consensus mechanism of blockchain. BVIB, based on VIB, approximates mutual information and utilizes machine learning to train the encoder and decoder in order to reduce the size of their required transmission data.}
The main contributions of our work are listed as follows:
\begin{itemize}
\item A new intelligent algorithm based on VIB and blockchain is proposed. We integrate VIB and blockchain to improve the network security while guarantee the decoding and compression performance of VIB. {Moreover, machine learning is integrated into our approach to train the encoding neural networks and decoding neural networks.} Note that this is the first integration of VIB and blockchain.
\item A new network framework is proposed. The framework splits the whole model into an encoder deployed on devices and a decoder deployed on {server}s. Thus, we can alleviate the computational load on user devices by moving a portion of the VIB network to {server}s. This can help mitigate the computation burden of user devices.
\item We build an experimental platform by integrating Python and C++ to verify the efficiency of our proposed BVIB approach, where Python is utilized for realizing the VIB algorithm and C++ is utilized to build a faster executing blockchain environment. The python and C++ program is linked by dynamic link librarytable (DLL), which is a common form of shared library.

\end{itemize}

The rest of this paper is organized as follows. Section \uppercase\expandafter{\romannumeral2} introduces the system model. Section \uppercase\expandafter{\romannumeral3} details the BVIB approach. Section \uppercase\expandafter{\romannumeral4} analyzes the simulation results. Section \uppercase\expandafter{\romannumeral5} concludes this paper.

\begin{figure}[]
	\centering
	\includegraphics[width=0.8\textwidth]{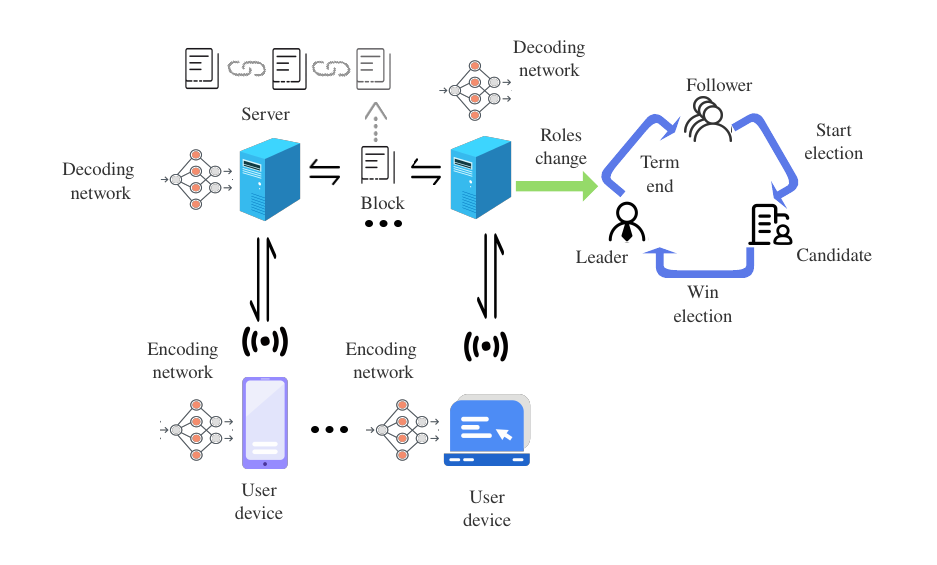}
	\caption{{System model.}}
	\label{fig1}
\end{figure}
\section{System Model}

In this section, we construct a network framework, which consists of various types of user devices (i.e., smartphones, computers, laptops) having insufficient computing resources and {server}s having sufficient computing resources. Each user device communicates with a {server}. Each {server} has a ledger to record data information and timestamp. {Server}s adopt the Raft consensus algorithm to guarantee network security \cite{hartlieb2022photo}. The Raft algorithm defines three roles (i.e., leader, follower and candidate) for {server}s. The leader is authorized exclusively to generate blocks and periodically broadcast control information to inform other {server}s of its existence. The follower records data information and sends the information to the leader. The candidate is an intermediate role, changing from a follower and waiting to become a leader. {Moreover, the consensus mechanism is divided into two parts. The first part is the election mechanism. Specifically, similar to\cite{10071537}, we first consider all followers as candidates in our work to simplify the process, which is different from the typical setup, i.e., a few followers who aspire to become leaders become candidates. Subsequently, each candidate casts a random vote, and the candidate with the highest number of votes becomes the new leader, while the remaining candidates return to being followers. The second part is block generation. Specifically, during the training, the followers send their ledgers to the leader. Then the leader collects them into a block. Later on, the followers update their ledgers upon the block content.}

Each user device and {server} deploys an encoder and a decoder (i.e.,  neural networks), respectively, where the neural networks of encoder and decoder are referred to as the encoding neural network and decoding neural network, respectively. Each user device carries datasets that include a training dataset and a testing dataset. The training and testing dataset are divided into $B$ batches. Denote $X={\{X_1,X_2,...,X_B\}}$, $Y={\{Y_1,Y_2,...,Y_B\}}$, $Z={\{Z_1,Z_2,...,Z_B\}}$ and $\hat{Y} = {\{\hat{Y}_1,\hat{Y}_2,...,\hat{Y}_B\}}$ as the data set, label set, encoding variable set and decoding variable set, respectively, where $X_i$, $Y_i$, $Z_i$ and $\hat{Y}_i$ $(i={1,2,...,B})$ are the data set, label set, encoding variable set and decoding variable set of the $i$th batch, respectively. For simplicity, we consider that the channel between each user device and {server} is ideal in this work and thus information can be transmitted among user devices and {server}s without error. The system model is illustrated as Fig.~\ref{fig1}.

\section{Blockchain-enabled Variational Information Bottleneck Approach}

In this section, we propose the BVIB approach to guarantee the network security. Specifically, the system first enters the training stage to train the encoding and decoding networks, then enters the testing stage to obtain the testing performance. In both training and testing stages, we introduce Denial of Service (DoS) attack where a malicious node sends significant amounts of useless requests to one node in the {server}s and user devices and makes the node paralyzed. The training and testing stages are introduced as follows. 
\subsection{Training stage}
At the beginning, the roles of all {server}s are set as candidates and {server}s adopt the election mechanism to elect a leader. Then the leader generates a genesis block to form a chain and sets a timer for a term to periodically broadcast its control information within the term. At the same time, the {server}s inform user devices to start training through $E$ epochs.

At the beginning of the $e$th epoch ($e={1,2,...,E}$), the parameters of the encoding and decoding networks are those updated at the end of the $(e-1)$th epoch. Note that when $e=1$, the parameters of encoding and decoding neural networks are randomly initialized. Then user devices input each one of the $B$ batches into the encoding neural network in order. For the $i$th ($i={1,2,...,B}$) batch, each user device inputs $X_i$ into the encoding neural network that outputs the encoding variable set $Z_i$. Let $z$ be any data belonging to $Z_i$. Similar to \cite{alemi2016deep}, we consider the probability of $z$ follows a Gaussian distribution, i.e.,
\begin{equation}
	p(z)=\mathcal{N}(z|\mu_i, \sigma^2_i),
	\qquad
	{\forall} z \in Z_i,
	\label{eq0}
\end{equation}
where $\mathcal{N}$ represents Gaussian distribution, ${\mu}_{i}$ and ${\sigma}^2_{i}$ are the mean and covariance matrix of encoding function, respectively. 

Based on the encoding assumption in \cite{alemi2016deep}, each user device can obtain ${\mu}_{i}$ and ${\sigma}^2_{i}$ satisfying \eqref{eq0} and send them to the corresponding {server}. Afterwards, each {server} adopts the reparameterization trick to generate reparameterized variable set $\hat{Z}$ based on the received mean and covariance. Let $\hat{z}$ be any data belonging to $\hat{Z_i}$, we have
\begin{equation}
\hat{z} = {\mu}_{i} + \epsilon{\sigma}^2_{i}, \qquad \forall \hat{z} \in \hat{Z},
\label{reparameterization}
\end{equation}
where $\epsilon \sim \mathcal{N}(0,1)$. According to \eqref{eq0} and \eqref{reparameterization}, we can find $z$ and $\hat{z}$ follow the same distribution.

Then each {server} inputs $\hat{Z}_i$ into the decoding network that outputs $\hat{Y}_i$. Afterward, each {server} calculates its lower bound of loss function to approximate the loss function. Since $z$ and $\hat{z}$ follow the same distribution, according to the constraint equation of VIB \cite{alemi2016deep}, each {server} calculates the mutual information bounds including the lower bound of the mutual information between $Z_i$ and $\hat{Y}_i$, denoted as $I(Z_i,\hat{Y}_i)_{min}$, and the upper bound of the mutual information between $Z_i$ and $X_i$, denoted as $I(Z_i,X_i)_{max}$, then calculates the lower bound of loss function based on the mutual information bounds, i.e.,

%\begin{algorithm}[h]
%	\caption{The training process of BVIB}
%	\label{alg:bVIB}
%	\begin{algorithmic}[1]
%		\REQUIRE Data $X$ from training dataset.
%		\STATE servers elect a leader.
%		
%		
%		\FOR{$e = 1$ to $E$}
%		\FOR{$i = 1$ to $B$}
%		\STATE Input $X_i$ into the encoding network, user devices obtain the encoding variable set $Z_i$. 
%		\STATE User devices obtain ${\mu}_{i}$ and ${\sigma}^2_{i}$.
%		\STATE User devices send ${\mu}_{i}$ and ${\sigma}^2_{i}$ to nearby servers.
%		\STATE servers calculate $\hat{Z_i}$ according to \eqref{reparameterization}.
%		\STATE servers input $\hat{Z_i}$ into the decoding network that outputs $\hat{Y}_i$.
%		\STATE servers calculate $\mathcal{L}_{min}$ according to \eqref{eq123}.
%		\STATE servers save $I(Z_i,X_i)_{max}$ and $I(Z_i,\hat{Y}_i)_{min}$ in the ledgers.
%		\STATE Adam algorithm is adopted to update the parameters of the encoding and decoding network.
%		\STATE The followers which are not attacked send their ledgers to the leader.
%		\STATE The leader collects the ledgers into a block and add it to the chain.
%		\ENDFOR
%		\ENDFOR
%	\end{algorithmic}
%\end{algorithm}

\begin{equation}
	\left\{
	\begin{aligned}
        &I(Z_i,\hat{Y}_i)_{min} = \sum_{n=1}^N\sum_{m=1}^M p(z_{n,i}|x_{m,i})\log_{{2}} q(\hat{y}_{m,i}|z_{n,i}), \\
		&I(Z_i,X_i)_{max} = \sum_{n=1}^N \sum_{m=1}^Mp(z_{n,i}|x_{m,i})\log_{{2}}\frac{p(z_{n,i}|x_{m,i})}{r(Z_i)}, \\		
        &\mathcal{L}_{min} =  I(Z_i,\hat{Y}_i)_{min}	- \beta I(Z_i,X_i)_{max}, \\
		&{\forall} z_{n,i} \in Z_i, \qquad \forall \hat{y}_{i} \in \hat{Y}_i,
	\end{aligned}
	\right.
	\label{eq123}
\end{equation}
where $M$ and $N$ are the number of data of $(X_i,\hat{Y}_i)$ and $Z_i$ in each batch, respectively, $x_{m,i}$ and $\hat{y}_{m,i}$ are the $m$th data of $X_i$ and $\hat{Y}_i$, respectively, $z_{n,i}$ is the $n$th data of $Z_i$, $p(z_{n,i}|x_{m,i})$ represents the conditional probability of $z_{n,i}$ given $x_{m,i}$, and $q(\hat{y}_{m,i}|z_{n,i})$ is a variational approximation to the conditional probability $p(\hat{y}_{m,i}|z_{n,i})$, $r(Z_i)$ is a variational approximation to the marginal of $Z_i$, $\beta$ is a Lagrange multiplier used to minimize the loss function. According to the principle of information bottleneck, we can make the decoding variable $\hat{Y}_i$ more aligned with the label information $Y_i$ by maximizing $I(Z_i,\hat{Y}_i)_{min}$, and achieve higher compression of the information by minimizing $I(Z_i,X_i)_{max}$ \cite{tishby2015deep}.

Then each user device and {server} adopts the Adam algorithm to update the parameters of the encoding and decoding neural networks, where {server}s send the gradients for the parameters of the first neural network layer to the user devices to update the parameters of the last neural network layer \cite{kingma2014adam}. After {server}s have sent the gradients, each {server} records the calculated $I(Z_i,X_i)_{max}$ and $I(Z_i,\hat{Y}_i)_{min}$ in its ledger.

{Then the followers which are not attacked send their ledgers to the leader. However, those followers which are attacked have paralyzed their own computing resources and are unable to send their ledgers. If the leader receives ledgers from less than half of the followers, the training stage is interrupted and then a new training stage is restarted. Otherwise, it collects all the received ledgers to generate a block, and adds it to the chain.} {The chain is deployed on and maintained by all the servers including the leader and followers. Thus, if an attacker wants to revise the content of a block of the chain, it needs to revise half the number of the servers at least which enhances the robustness of the network.} Then the leader broadcasts the new block to all followers to update their ledgers. After that the {server}s including followers and the leader will inform the user devices to input the next batch into the encoding network to continue training. The above training process is repeated until the batch number $i$ reaches $B$. Then the next epoch starts. The training process continues until either $\mathcal{L}_{min}$ converges or the epoch number $e$ reaches $E$. 
%The training process is illustrated in Algorithm \ref{alg:bVIB}.

%Note that during the training stage, if the leader is attacked or the timer reaches a term, which resets the timer and periodically broadcasts the control information, the election mechanism is executed to elect a new leader.

{Note that during the training stage, if the leader is attacked and paralyzed. It is unable to periodically broadcast the control information. Moreover, the followers cannot receive the control information, the election mechanism is executed to elect a new leader. Furthermore, if the timer reaches a term, the same steps repeat. }

\subsection{Testing stage}

Compared with the training stage, the testing stage omits the parameters updating process. The system adopts the trained parameters and inputs the testing dataset to obtain the testing $\hat{Y}_i$. During the testing stage, $I(Z_B,X_B)_{max}$ and $I(Z_B,\hat{Y_B})_{min}$ of each epoch are recorded and the accuracy of each epoch is calculated as $(1-\frac{1}{B}\frac{1}{M}\sum_{i=1}^{B}\sum^M_{m=1}\hat{y}_{m,i}\oplus y_{m,i})\times100\%$. In the following context, $I(Z_B,X_B)_{max}$ and $I(Z_B,\hat{Y_B})_{min}$ are denoted as $I(Z,X)_{max}$ and $I(Z,\hat{Y})_{min}$ for simplicity. After the testing stage, the time and central processing unit (CPU) cycles cost are recorded. Moreover, the testing average accuracy is calculated as the average of all epochs'.

%which is denoted as $I(Z,X)_{max}$ for simplicity,

%testing results including   each {server} use $\hat{Y}_i$ and $Y_i$ to calculate the accuracy ($(1-\frac{1}{M}\sum^M_{m=1}\hat{y}_{m,i}\oplus y_{m,i})\times100\%$). %($(1-\frac{1}{B}\frac{1}{M}\sum_{i=1}^{B}\sum^M_{m=1}\hat{y}_{m,i}\oplus y_{m,i})\times100\%$)

\section{Performance Evaluation}

\begin{figure*}[] %通栏
	\begin{minipage}[t]{0.33\linewidth} %调节两个子图左右间距
		\centering
		\includegraphics[width=2.5in, height=1.9in]{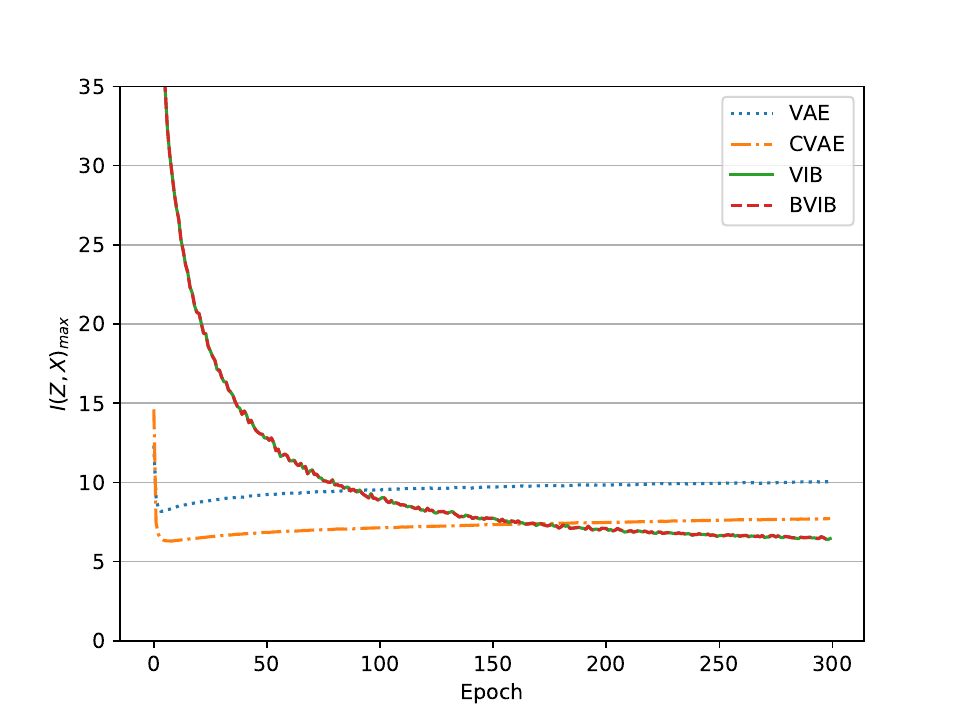} %调节单个子图大小
		\caption{$I(Z,X)_{max}$ under different approaches.} %子图下标题
		\label{fig2} %引用标签
	\end{minipage}%
	\begin{minipage}[t]{0.33\linewidth}
		\centering
		\includegraphics[width=2.5in, height=1.9in]{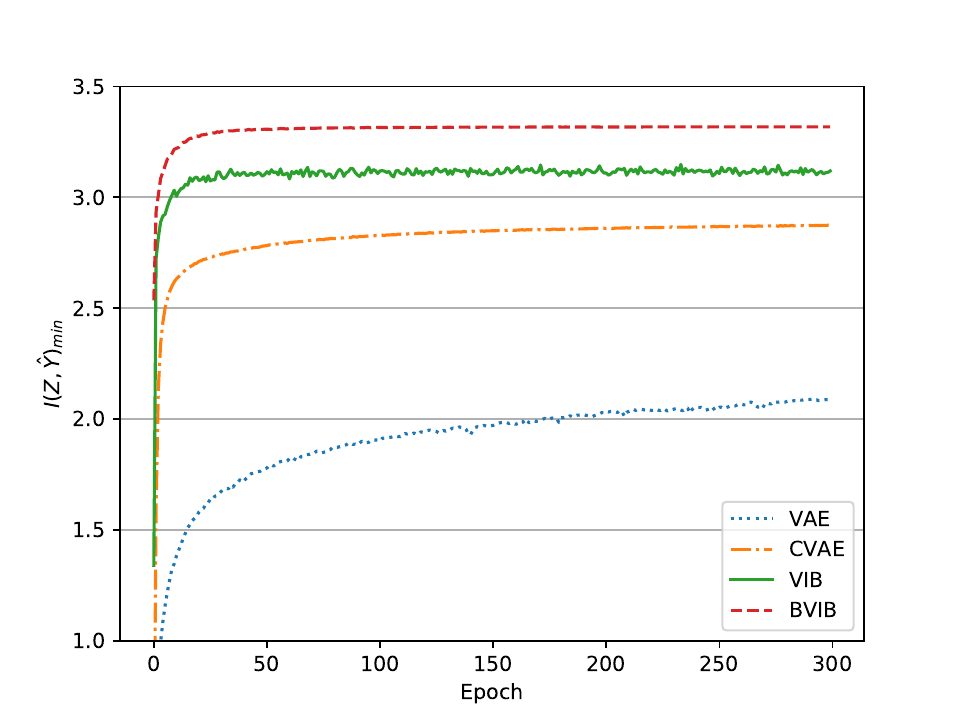}
		\caption{$I(Z,\hat{Y})_{min}$ under different approaches.}
		\label{fig3}
	\end{minipage}%
	\begin{minipage}[t]{0.33\linewidth}
		\centering
		\includegraphics[width=2.5in, height=1.9in]{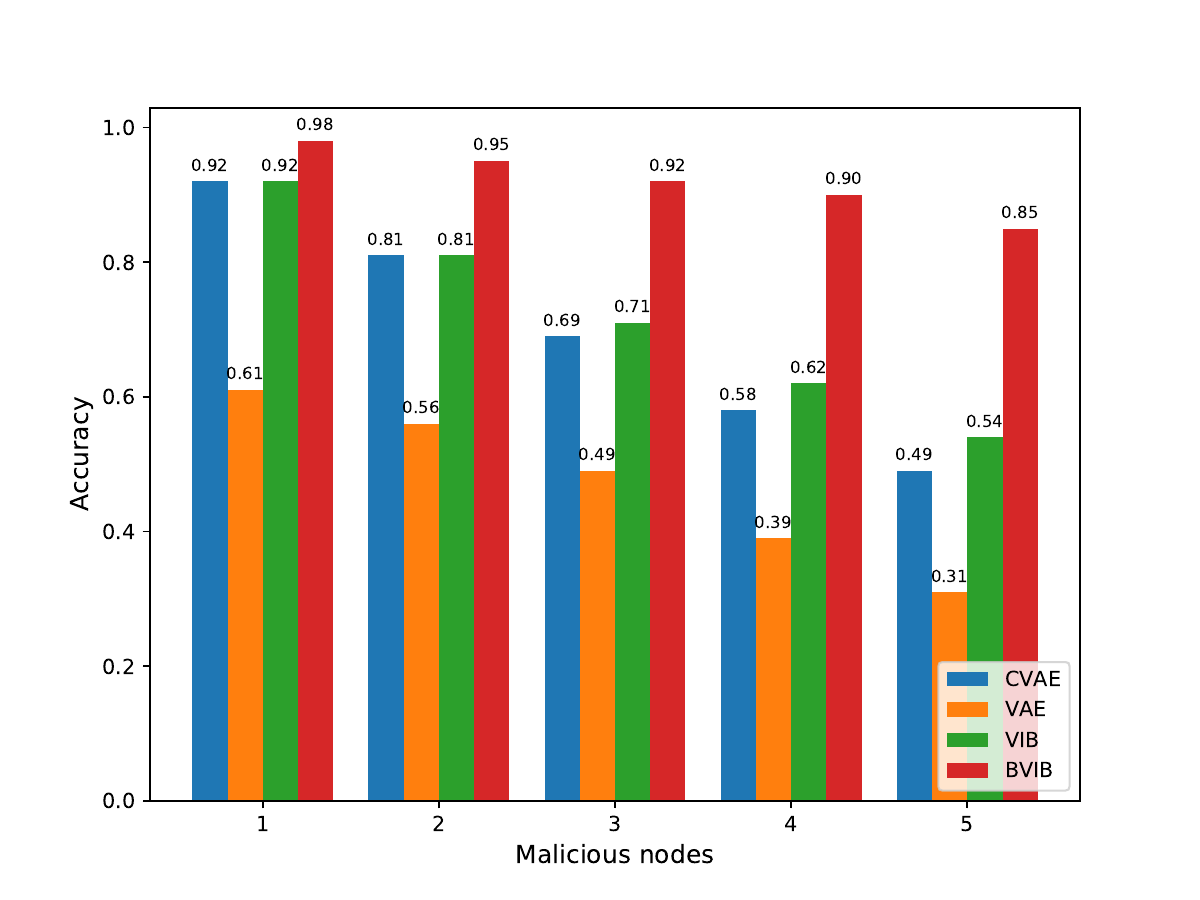}
		\caption{{Accuracy under different number of malicious nodes and approaches.}}
		\label{fig4}
	\end{minipage}
\end{figure*}

\begin{table}[htbp]
\caption{Parameter Settings.}
	\begin{center}
		\begin{tabular}{|c|c|c|}
			\hline
			\textbf{Description}&{\textbf{Value}} \\
			\hline
			$E$ & 300\\
			\hline
			$B$ & 200 \\
%			\hline
%			The number of dataset divided & 200 \\
			\hline
			Learning rate & 0.001 \\
			\hline
			Term & 10min \\
%			\hline
%			Layers of model & 5 \\
			\hline
			Maximum frequency of CPU & 5.20GHz\\
			\hline
			Number of user device & 10\\
			\hline
			Number of {server}s & 10\\
			\hline
			Activate function & ReLU \\
			\hline
			Optimizer & Adam \\
%			\hline
%			Layers of encoder & 3 \\
			\hline
			Structure of encoder & 784-1024-512 \\
			\hline
			Structure of decoder & 512-784-10 \\

			\hline
		\end{tabular}
		\label{tab1}
	\end{center}
\end{table}

In this section, we build an experimental platform and conduct simulation experiments to verify the efficiency of our proposed BVIB approach through comparing with the baseline approaches. {The experimental platform is built by integrating Python and C++, where Python is utilized for realizing the VIB algorithm and C++ is utilized to build a faster executing blockchain environment. The python and C++ program are linked by DLL. The modified national institute of standards and technology (MNIST) dataset which contains 60,000 training data and 10,000 test data is adopted in the experiments\cite{10091897}. The detailed parameter settings are listed in Table \ref{tab1}.
%	\footnote{The dataset is available at http://yann.lecun.com/exdb/mnist}\cite{10091897}.} The detailed parameter settings are listed in Table \ref{tab1}.

The baseline approaches are introduced as follows.

(1) Variational autoencoder (VAE): VAE is an unsupervised learning method aiming to learn the compact representation of input data. As VAE is a classical algorithm in deep learning, we choose it as a comparable algorithm in our work.

(2) Conditional variational autoencoder (CVAE): CVAE is an extension of the VAE that allows for the incorporation of conditional information during the learning process to generate data.

(3) VIB: VIB is the first trial that applies variational inference in information bottleneck. It enables information bottleneck to process complex data and make mutual information easy to be calculated.

According to the structure of encoder and decoder, the number of neurons on each user device and {server} are 2320 and 1306, respectively. In the following experiments, we set the number of neurons on each user device for VAE, CVAE and VIB to be the same as the total neurons on both user devices and {server}s for BVIB, i.e., 3626. Each following result is the average value of 10 experiments.

%\begin{figure*}[htbp] %通栏
%	\begin{minipage}[t]{0.33\linewidth} %调节两个子图左右间距
%		\centering
%		\includegraphics[width=2.5in, height=1.9in]{izx3.pdf} %调节单个子图大小
%		\caption{$I(Z,X)_{max}$ under different approaches.} %子图下标题
%		\label{fig2} %引用标签
%	\end{minipage}%
%	\begin{minipage}[t]{0.33\linewidth}
%		\centering
%		\includegraphics[width=2.5in, height=1.9in]{izy3.pdf}
%		\caption{$I(Z,\hat{Y})_{min}$ under different approaches.}
%		\label{fig3}
%	\end{minipage}%
%	\begin{minipage}[t]{0.33\linewidth}
%		\centering
%		\includegraphics[width=2.5in, height=1.9in]{an1.pdf}
%		\caption{Accuracy under attack.}
%		\label{fig4}
%	\end{minipage}
%\end{figure*}

Fig.~\ref{fig2} illustrates $I(Z,X)_{max}$ with the increasing epoch number under the four approaches. {We can see that VIB and BVIB have a smooth decreasing trend, and after epoch 240, they eventually drop to a level even lower than those of VAE and CVAE by 40\% and 20\%.} This is because the loss function of BVIB and VIB effectively focuses on optimizing mutual information. In contrast, VAE and CVAE only calculate cross-entropy between input and output in each epoch, without considering mutual information. {In addition, we can see that the values of $I(Z,X)_{max}$ under VIB and BVIB are almost the same}, which demonstrates that BVIB is capable of achieving nearly the same compression performance as VIB. Moreover, CVAE converges to a small $I(Z,X)_{max}$ compared to VAE for the reason that CVAE can use conditional information to obtain better compression capacity.

Fig.~\ref{fig3} illustrates $I(Z,\hat{Y})_{min}$ with the increasing epoch number under the four approaches. {We can see that VIB and BVIB converge to a larger value than those of VAE and CVAE by 22\% and 43\%, because the loss function of VIB and BVIB aims to maximize $I(Z,\hat{Y})_{min}$. In addition, we can see that $I(Z,\hat{Y})_{min}$ of BVIB is larger than that of VIB by 7\% for the reason that BVIB is robust against attack.} Moreover, it is seen that the converged value under VAE is smaller than that under CVAE, because VAE is unsupervised learning while CVAE is supervised learning, and the training performance of unsupervised learning is inferior to that of supervised learning.

\begin{table}[]
	\caption{Comparison of resource consumption}
	\begin{center}
		\begin{tabular}{|c|c|c|c|}
			\hline
			\textbf{Device}&{\textbf{Neurons}}&{\textbf{Time}}&{\textbf{CPU cycles}} \\
			\hline
			VAE(single device) & 3626& 83.1s&254.73G \\
			\hline
			CVAE(single device) & 3626& 83.3s&261.12G \\
			\hline
			VIB(single device) & 3626& 83.2s&257.92G \\
			\hline
			BVIB(user device) & 2320& 53.1s&164.61G  \\
			\hline
			BVIB({server}) & 1306& 32.1s&99.51G  \\
			\hline
			{BVIB(user device + server)} & {3626} & {85.2s} & {264.12G}\\
			\hline
			%			\multicolumn{4}{l}{$^{\mathrm{a}}$Sample of a Table footnote.}
		\end{tabular}
		\label{tab2}
	\end{center}
\end{table}

Table \ref{tab2} shows the resource consumption of VAE, CVAE, VIB and BVIB. {We can find that the time and CPU cycles cost of either a user device or a {server} in the proposed approach are smaller than that of other baseline approaches by 36\%.} This is because each user device in BVIB has fewer neurons. {By placing the encoder networks on user devices and decoder networks on {server}s, the total time and CPU cycles cost of BVIB are slightly more than others. However, BVIB significantly reduces the computation demand of a user device.}

Fig.~\ref{fig4} compares the accuracy of VAE, CVAE, VIB and BVIB under different numbers of malicious nodes. We can see that the accuracy of the four approaches is degraded when the number of malicious nodes increases, because more nodes are attacked and become paralyzed. {Moreover, when the number of malicious nodes reaches 5, BVIB has a much higher accuracy than VAE, CVAE and VIB by 174\%, 73\% and 57\%, respectively.} This is because BVIB approach can elect a new leader when the old one is attacked, which prevents the malicious nodes from attack and guarantees the network security.

\section{Conclusions}
In this paper, we have proposed the BVIB approach to alleviate the computation burden and guarantee network security in IoT networks. Extensive simulations have been conducted to demonstrate the performance of BVIB. {The conclusions are drawn as follows: By combining the VIB and blockchain, BVIB can guarantee the IoT network security while guaranteeing the performance of VIB;	Our proposed framework splits the model into an encoder deployed on devices and a decoder deployed on {server}s. Therefore, the computing burden of user devices can be relieved.}

%\begin{itemize}
%\item By combining the VIB and blockchain, BVIB can guarantee the IoT network security while guaranteeing the performance of VIB.	
%\item Our proposed framework splits the model into an encoder deployed on devices and a decoder deployed on {server}s. Therefore, the computing burden of user devices can be relieved.
%	
%\end{itemize}

\bibliographystyle{IEEEtran.bst}
%\balance
\bibliography{ref}

\begin{thebibliography}{10}
\providecommand{\url}[1]{#1}
\csname url@samestyle\endcsname
\providecommand{\newblock}{\relax}
\providecommand{\bibinfo}[2]{#2}
\providecommand{\BIBentrySTDinterwordspacing}{\spaceskip=0pt\relax}
\providecommand{\BIBentryALTinterwordstretchfactor}{4}
\providecommand{\BIBentryALTinterwordspacing}{\spaceskip=\fontdimen2\font plus
\BIBentryALTinterwordstretchfactor\fontdimen3\font minus
  \fontdimen4\font\relax}
\providecommand{\BIBforeignlanguage}[2]{{%
\expandafter\ifx\csname l@#1\endcsname\relax
\typeout{** WARNING: IEEEtran.bst: No hyphenation pattern has been}%
\typeout{** loaded for the language `#1'. Using the pattern for}%
\typeout{** the default language instead.}%
\else
\language=\csname l@#1\endcsname
\fi
#2}}
\providecommand{\BIBdecl}{\relax}
\BIBdecl

\bibitem{9064704}
P.~Chanak and I.~Banerjee, ``{Congestion Free Routing Mechanism for IoT-Enabled
  Wireless Sensor Networks for Smart Healthcare Applications},'' \emph{IEEE
  Transactions on Consumer Electronics}, vol.~66, no.~3, pp. 223--232, 2020.

\bibitem{9785856}
G.~Kaur, P.~Chanak, and M.~Bhattacharya, ``{A Green Hybrid Congestion
  Management Scheme for IoT-Enabled WSNs},'' \emph{IEEE Transactions on Green
  Communications and Networking}, vol.~6, no.~4, pp. 2144--2155, 2022.

\bibitem{tishby2000information}
N.~Tishby, F.~C. Pereira, and W.~Bialek, ``{The Information Bottleneck
  Method},'' \emph{arXiv preprint physics/0004057}, 2000.

\bibitem{10313285}
Z.~Li, R.~She, P.~Fan, C.~Peng, and K.~B. Letaief, ``{Learning Channel Capacity
  with Neural Mutual Information Estimator Based on Message Importance
  Measure},'' \emph{IEEE Transactions on Communications}, 2023.

\bibitem{tishby2015deep}
N.~Tishby and N.~Zaslavsky, ``{Deep Learning and The Information Bottleneck
  Principle},'' in \emph{2015 IEEE Information Theory Workshop (ITW)},
  Jerusalem, Israel, 2015, pp. 1--5.

\bibitem{alemi2016deep}
A.~A. Alemi, I.~Fischer, J.~V. Dillon, and K.~Murphy, ``{Deep Variational
  Information Bottleneck},'' \emph{arXiv preprint arXiv:1612.00410}, 2016.

\bibitem{wu2019trajectory}
Q.~Wu, H.~Liu, C.~Zhang, Q.~Fan, Z.~Li, and K.~Wang, ``{Trajectory Protection
  Schemes Based on a Gravity Mobility Model in IoT},'' \emph{Electronics},
  vol.~8, no.~2, 2019.

\bibitem{9868080}
S.~Banaeian~Far and A.~Imani~Rad, ``{PP-DENT: A Privacy-Preserving Framework
  for Blockchain-Based Mobile/Roaming Transactions},'' \emph{IEEE Networking
  Letters}, vol.~4, no.~4, pp. 204--207, 2022.

\bibitem{10024370}
K.~Saadat, N.~Wang, and R.~Tafazolli, ``{AI-Enabled Blockchain Consensus Node
  Selection in Cluster-Based Vehicular Networks},'' \emph{IEEE Networking
  Letters}, vol.~5, no.~2, pp. 115--119, 2023.

\bibitem{10158495}
C.~Wang, S.~Du, W.~Sun, and D.~Fan, ``{Self-Supervised Learning for
  High-Resolution Remote Sensing Images Change Detection With Variational
  Information Bottleneck},'' \emph{IEEE Journal of Selected Topics in Applied
  Earth Observations and Remote Sensing}, vol.~16, pp. 5849--5866, 2023.

\bibitem{9813696}
M.~P. Uddin, Y.~Xiang, X.~Lu, J.~Yearwood, and L.~Gao, ``{Federated Learning
  via Disentangled Information Bottleneck},'' \emph{IEEE Transactions on
  Services Computing}, vol.~16, no.~3, pp. 1874--1889, 2023.

\bibitem{li2023task}
H.~Li, C.~Zhu, Y.~Zhang, Y.~Sun, Z.~Shui, W.~Kuang, S.~Zheng, and L.~Yang,
  ``{Task-specific Fine-tuning via Variational Information Bottleneck for
  Weakly-supervised Pathology Whole Slide Image Classification},'' in
  \emph{Proceedings of the IEEE/CVF Conference on Computer Vision and Pattern
  Recognition}, Vancouver, BC, Canada, 2023, pp. 7454--7463.

\bibitem{wei2021secure}
X.~Wei, Y.~Yan, S.~Guo, X.~Qiu, and F.~Qi, ``{Secure Data Sharing:
  Blockchain-Enabled Data Access Control Framework for IoT},'' \emph{IEEE
  Internet of Things Journal}, vol.~9, no.~11, pp. 8143--8153, 2021.

\bibitem{nguyen2021secure}
D.~C. Nguyen, P.~N. Pathirana, M.~Ding, and A.~Seneviratne, ``{Secure
  Computation Offloading in Blockchain Based IoT Networks With Deep
  Reinforcement Learning},'' \emph{IEEE Transactions on Network Science and
  Engineering}, vol.~8, no.~4, pp. 3192--3208, 2021.

\bibitem{10131971}
X.~Lan, X.~Tang, R.~Zhang, W.~Lin, and Z.~Han, ``{UAV-Assisted Computation
  Offloading Toward Energy-Efficient Blockchain Operations in Internet of
  Things},'' \emph{IEEE Wireless Communications Letters}, vol.~12, no.~8, pp.
  1469--1473, 2023.

\bibitem{hartlieb2022photo}
M.~Hartlieb, ``{Photo-Iniferter RAFT Polymerization},'' \emph{Macromolecular
  Rapid Communications}, vol.~43, no.~1, p. 2100514, 2022.

\bibitem{10071537}
Y.~Li, Y.~Fan, L.~Zhang, and J.~Crowcroft, ``{RAFT Consensus Reliability in
  Wireless Networks: Probabilistic Analysis},'' \emph{IEEE Internet of Things
  Journal}, vol.~10, no.~14, pp. 12\,839--12\,853, 2023.

\bibitem{kingma2014adam}
D.~P. Kingma and J.~Ba, ``{Adam: A Method for Stochastic Optimization},''
  \emph{arXiv preprint arXiv:1412.6980}, 2014.

\bibitem{10091897}
Q.~Wu, X.~Wang, Q.~Fan, P.~Fan, C.~Zhang, and Z.~Li, ``{High Stable And
  Accurate Vehicle Selection Scheme Based on Federated Edge Learning in
  Vehicular Networks},'' \emph{China Communications}, vol.~20, no.~3, pp.
  1--17, 2023.

\end{thebibliography}

\vfill

\end{document}